\documentclass[12pt,preprint]{aastex}

\usepackage{graphicx}
\usepackage{epstopdf}

\newcommand{\ms}{\mathrm{m\,s^{-1}}}
\newcommand{\myemail}{christoph.bergmann@pg.canterbury.ac.nz}

\shorttitle{Searching for Earth-mass planets around $\alpha$ Centauri}
\shortauthors{Bergmann et al.}

\begin{document}

\title{Searching for Earth-mass planets around\\ $\alpha$ Centauri: precise radial velocities from contaminated spectra}

\author{Christoph Bergmann}
\affil{Department of Physics \& Astronomy, University of Canterbury, Private Bag 4800, Christchurch 8140, New Zealand}		
\email{\myemail}
\author{Michael Endl}
\affil{McDonald Observatory, The University of Texas at Austin, Austin, TX 78712}
\author{John B. Hearnshaw}
\affil{Department of Physics \& Astronomy, University of Canterbury, Private Bag 4800, Christchurch 8140, New Zealand}
\author{Robert A. Wittenmyer}
\affil{Department of Astrophysics and Optics, School of Physics, University of New South Wales, Sydney 2052, Australia}
\author{Duncan J. Wright}
\affil{Department of Astrophysics and Optics, School of Physics, University of New South Wales, Sydney 2052, Australia}

\begin{abstract}
This work is part of an ongoing project which aims to detect terrestrial planets in our neighbouring star system $\alpha$ Centauri using the Doppler method. Owing to the small angular separation between the two components of the $\alpha$ Cen AB binary system, the observations will to some extent be contaminated with light coming from the other star.

We are accurately determining the amount of contamination for every observation by measuring the relative strengths of the H-$\alpha$ and NaD lines. Furthermore, we have developed a modified version of a well established Doppler code that is modelling the observations using two stellar templates simultaneously. With this method we can significantly reduce the scatter of the radial velocity measurements due to spectral cross-contamination and hence increase our chances of detecting the tiny signature caused by potential Earth-mass planets. After correcting for the contamination we achieve radial velocity precision of $\sim 2.5\,\ms$ for a given night of observations.

We have also applied this new Doppler code to four southern double-lined spectroscopic binary systems (HR159, HR913, HR7578, HD181958) and have successfully recovered radial velocities for both components simultaneously.\\
\end{abstract}


\section{Introduction}
Since the discovery of the first extrasolar planet around a main sequence star \citep{firstplanet} about two decades ago, almost 2000 extrasolar planets have been discovered, owing in part to the vast success of NASA's \textit{Kepler} mission \citep{kepler}. The long desired goal of finding an Earth-mass planet in the habitable zone (HZ) of a Sun-like star, which is often referred to as the ``holy grail'' of planet hunting, is now within reach, as the recent discovery of Kepler-186f proves \citep{kepler186}. Even more intriguing, if such a planet is found orbiting a nearby Sun-like star, the potential for extensive follow-up observations will be enormous.

\citet{dumusque2012} announced the discovery of an Earth-mass planet in a 3.24-d orbit around $\alpha$ Cen B using 459 precise radial velocity (RV) measurements taken with the HARPS spectrograph \citep{harps_mayor2003}. Although this would place the planet far from the star's HZ \citep{kaltenegger2013}, it remains an intriguing claim, as it shows that the tiny RV signals from Earth-mass planets may be detectable even if they are buried in the RV noise. However, \cite{hatzes2013} re-analysed the HARPS data and his findings did not confirm the existence of the planet. It is, therefore, important that the reality of the purported planet be double-checked with an independent data set.

We have undertaken an observationally intensive campaign to detect Earth-mass planets in our neighbouring star system, $\alpha$ Centauri, including those which may lie within the respective habitable zone of either star. Over the last five years we have taken about $26\,500$ spectra of $\alpha$ Cen A and about $20\,000$ spectra of $\alpha$ Cen B with the HERCULES spectrograph \citep{HERCULES} attached to the 1-m McLellan telescope at Mt John University Observatory (MJUO), New Zealand, to measure radial velocities of both stars. We have also obtained between 100 and 200 spectra of each of four moderately bright double-lined spectroscopic binaries (SB2s). Our observational campaign was described in much more detail by \citet{jbh_narit}, \citet{rob2014} and \citet{mike_astrobio}.

Here we present our approach to overcoming the problem of spectral cross-contamination in the spectra of the $\alpha$ Cen AB binary system. When we started observing in 2009 the angular separation between the two components was about $8''$, but unfortunately it is steadily decreasing (only $\sim 4.5''$ at present, decreasing to a local minimum of $\sim 4''$ in Dec 2015). The typical seeing conditions at MJUO are $\sim 2.5''$ and the optical fibre connecting the telescope to the spectrograph subtends an angle of $\sim 4.5''$ on the sky. Therefore, when taking a spectrum of either $\alpha$ Cen A or B, some light from the other star will enter the fibre as well, especially for observations taken in poorer seeing conditions. In order to alleviate that problem we are using a $3''$ diameter pinhole, but that does not solve the problem entirely. This contamination results in a net velocity shift as the two stars have a radial velocity difference due to their binary orbit (currently about $2.8\,\mathrm{km\,s^{-1}}$, increasing by $\sim 1.2\,\mathrm{m\,s^{-1}\,d^{-1}}$). In order to detect the tiny Doppler wobble of the stars due to a potential terrestrial planet $(\lesssim 1\,\mathrm{m\,s^{-1}})$, it is imperative that the true uncontaminated velocities can be obtained.

\section{Methods}
The basic idea is to treat $\alpha$ Centauri as a double-lined spectroscopic binary system with a very large and variable flux ratio. We developed a routine that uses the line index ratio (a measure of relative line strengths) of the $\mathrm{H}$-$\alpha$ and NaD lines to quantify the amount of contamination, i.e. to determine the flux ratio of a given observation. The amount of contamination mainly depends on the seeing conditions and on imperfect telescope guiding. From that we can in principle derive the corresponding uncontaminated velocities given that the correlation between the line index ratio and the corresponding RV shift is known.

However, more reliable results can be obtained by using the flux ratio as well as the difference in RV due to the binary orbit \citep{orbit} as input parameters to a modified version of a well established Doppler code first described by \citet{austral}. This code now uses two stellar templates simultaneously to model the observations. We use the simultaneous iodine technique for wavelength reference and reconstruction of the instrumental profile. The full model can be written in general terms as\\
\begin{equation}
g(\lambda) = \kappa\,\Big[\Big(f_1\left(\lambda + \Delta\lambda_1\right) + \epsilon f_2\left(\lambda + \Delta\lambda_2\right)\Big)\,f_{I_2}\left(\lambda\right)\Big]\;\otimes\;\phi \qquad,
\label{eq:sb2model}\\
\end{equation}\\where $\kappa$ is a normalization factor, $f_1$ and $f_2$ are the stellar templates, $\Delta\lambda_1$ and $\Delta\lambda_2$ are their respective Doppler shifts, $f_{I_2}$ is the iodine template, $\epsilon\,(\ll 1)$ is the amount of contamination, $\phi$ is the instrumental profile and $\otimes$ denotes a convolution.

Note that generally $\epsilon$ is different for every observation of $\alpha$ Cen A or B as the amount of contamination changes, whereas the flux ratio is constant for the SB2 systems, unless they are eclipsing or intrinsically variable stars. On the other hand both shift parameters are freely modelled for the SB2 systems, but, because the contaminating star usually does not contribute a lot of light, the optimum value of its shift parameter is not very well defined for observations of $\alpha$ Cen. It is therefore more effective to couple it to the shift parameter of the main target using the known binary orbit and also the differences in convective blueshift and gravitational redshift between the two stars \citep[from][]{orbit}.

\section{Results}
Because the passbands used for the calculation of the line index ratio $R$ contain several telluric water lines, and because several other telluric lines move in and out of these passbands over the course of a year as the Earth orbits the Sun, the exact value of the line index ratio $R$ depends on the water vapour column density and the barycentric correction for a given observation. After correcting for these two effects, and after subtracting the binary orbit, we found a linear correlation between the line index ratio $R$ and the corresponding spurious shift in the uncorrected RVs, both from synthetic observations and from real observations (Fig. 1). The corrected line index ratio can therefore be used as a reliable measure of the amount of contamination.

\begin{figure}[t]
	\centering
	  \includegraphics[width=0.49\textwidth]{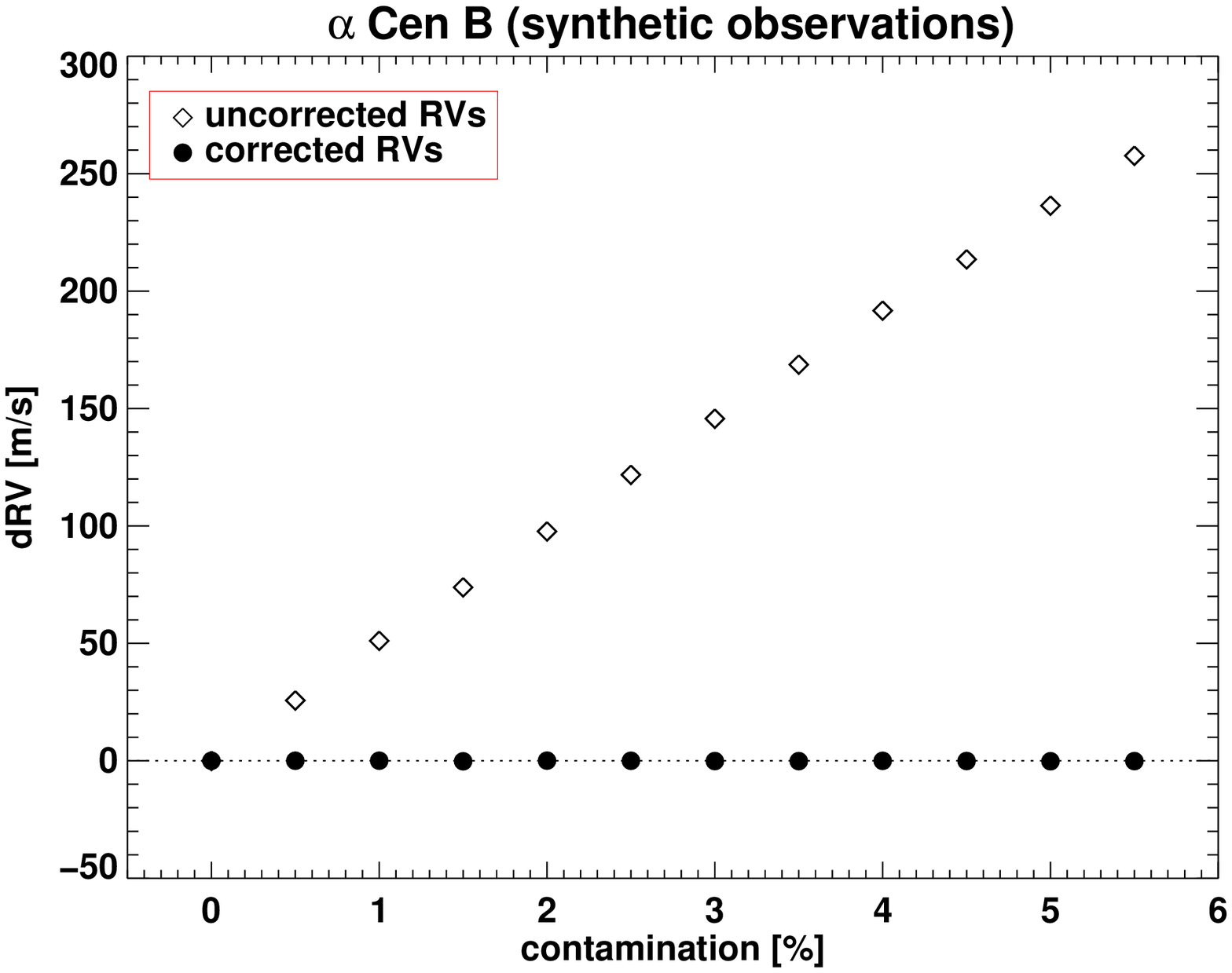}
		\includegraphics[width=0.49\textwidth]{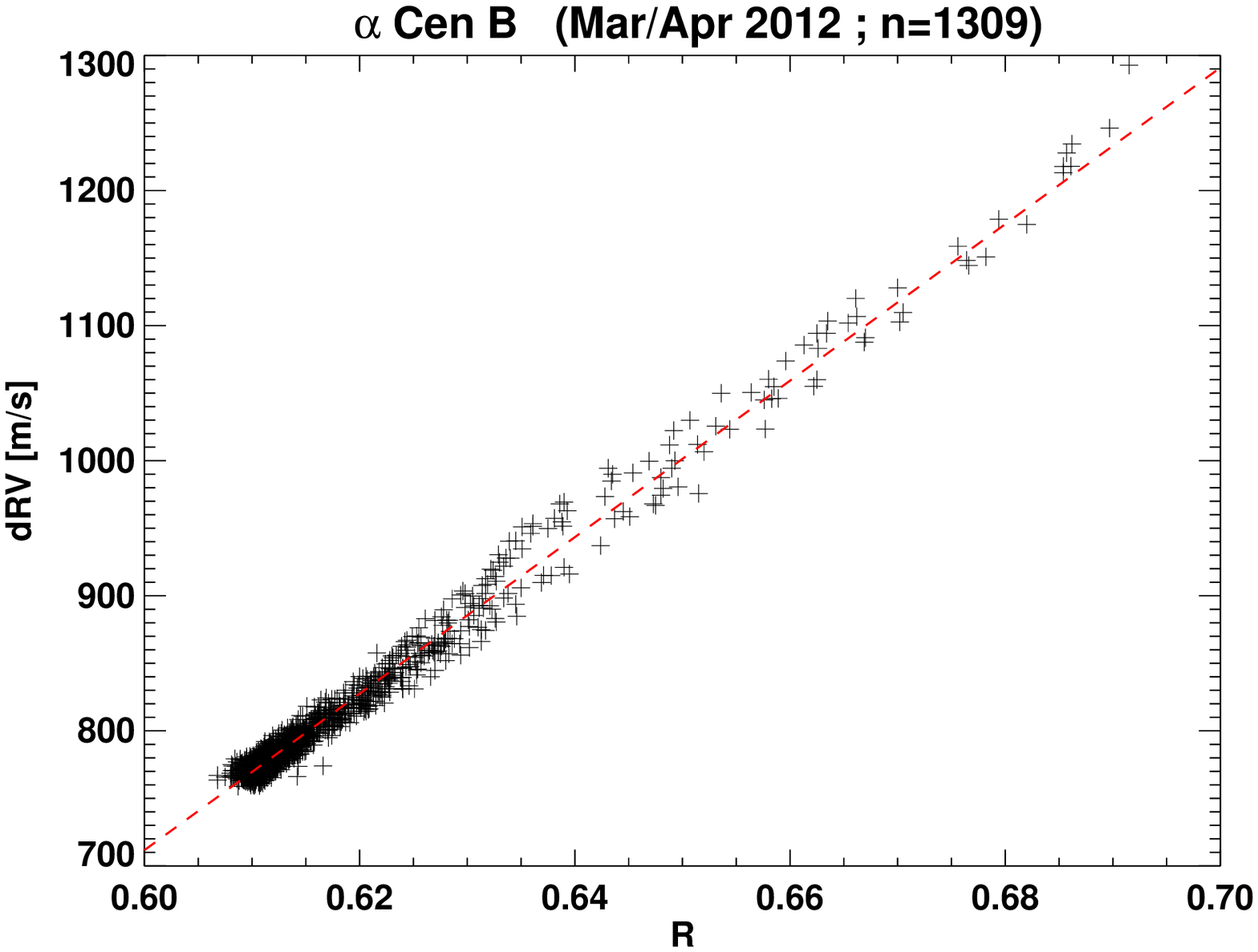}
	\caption{\textit{left panel:} Linear correlation between the amount of contamination and the effective shift in the measured RV if the contamination is not accounted for (diamonds). When accounting for the contamination (dots), the RV is independent of the amount of contamination. All data points were produced using synthetic observations of $\alpha$ Cen B with varying amounts of contamination from $\alpha$ Cen A, with $\Delta \mathrm{RV} = 2\,\mathrm{km\,s^{-1}}$.	\textit{right panel:} Linear correlation between line index ratio $R$ and spurious shift in the uncorrected RVs for observations of $\alpha$ Cen B from an observing run in March/April 2012 (right panel). Greater values of $R$ mean more contamination from $\alpha$ Cen A.}
	\label{fig:RV_vs_R}
\end{figure}

The true uncontaminated velocities were obtained using the modified Doppler code described above. Figure 2 shows both the uncorrected (contaminated) velocities and the corrected (uncontaminated) velocities for 1307 observations of $\alpha$ Cen B taken during an observing run in March/April 2012. The RMS for one night is typically $2.5\,\ms$ after correcting for the contamination, while it is typically $20\,\ms$ if the contamination is unaccounted for and can even exceed $100\,\ms$ for some heavily affected nights. The RMS for all 1307 observations is reduced from $70.2\,\ms$ to $4.8\,\ms$, which is a decrease by a factor of almost 15. Discarding the most heavily affected $20\%$ of these observations, the RMS is even further reduced to $3.4\,\ms$. Any residual RV variations are then due to intrinsic stellar activity and the presence of any potential planets. We have performed a period analysis of these 1307 corrected RVs but did not find any significant peaks in the periodogram. We are currently in the process of reducing all $\sim 47\,000$ observations of the two stars with the method described above and expect that we can scrutinize the reality of the putative planet in a 3.24-d orbit around $\alpha$ Cen B once we have a full set of corrected RVs.

\begin{figure}[t]
	\centering
		\includegraphics[width=0.80\textwidth]{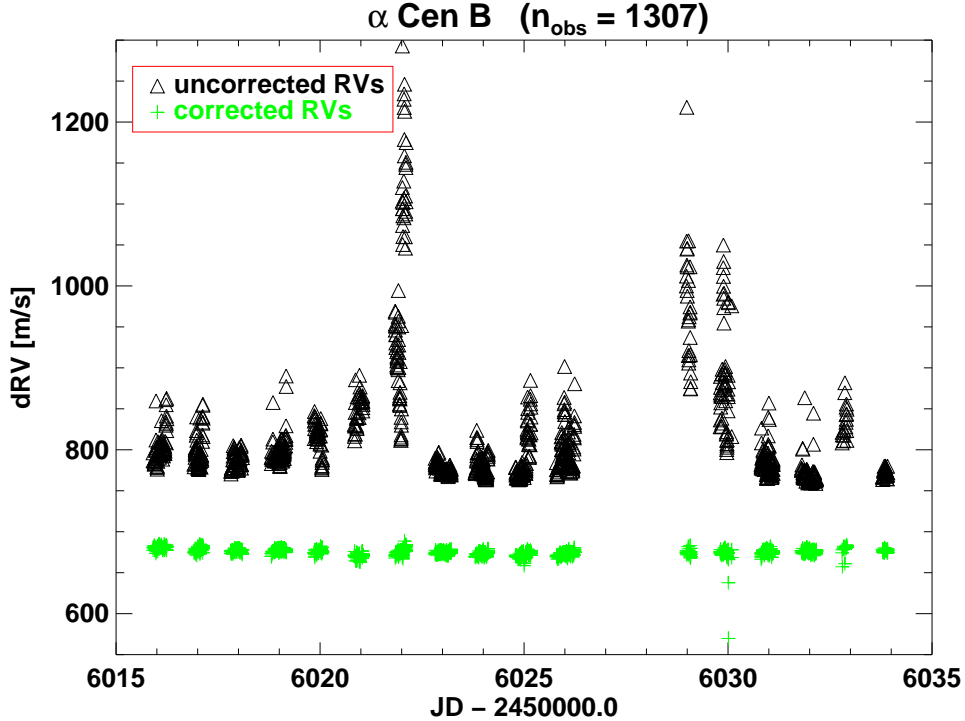}
	\caption{Corrected and uncorrected radial velocities for 1307 spectra of $\alpha$ Cen B taken over 17 nights in March/April 2012. In both cases the binary orbit has already been subtracted. Note that these are relative velocities (relative to the template) and hence the zero point for the corrected velocities has been arbitrarily shifted for the sake of clarity.}
\end{figure}

This modified Doppler code can also be applied to other SB2 systems. As a proof of method we have successfully recovered precise RVs for both components of four southern SB2 systems (HR159, HR913, HR7578, HD181958). Radial velocity measurements for HR7578, which consists of a pair of almost equal K3 dwarfs, are shown in Fig.~3 as an example. The Keplerian orbit was determined using the RVLIN package (Wright \& Howard~2009, Wright \& Howard~2013) and the corresponding uncertainties in the orbital parameters were estimated via a bootstrap algorithm \citep{boottran}. Our best-fit orbital solution is in excellent agreement with the one found by \citet{hd188088} and of much higher accuracy. 

\begin{figure}[t]
	\centering
		\includegraphics[width=0.80\textwidth]{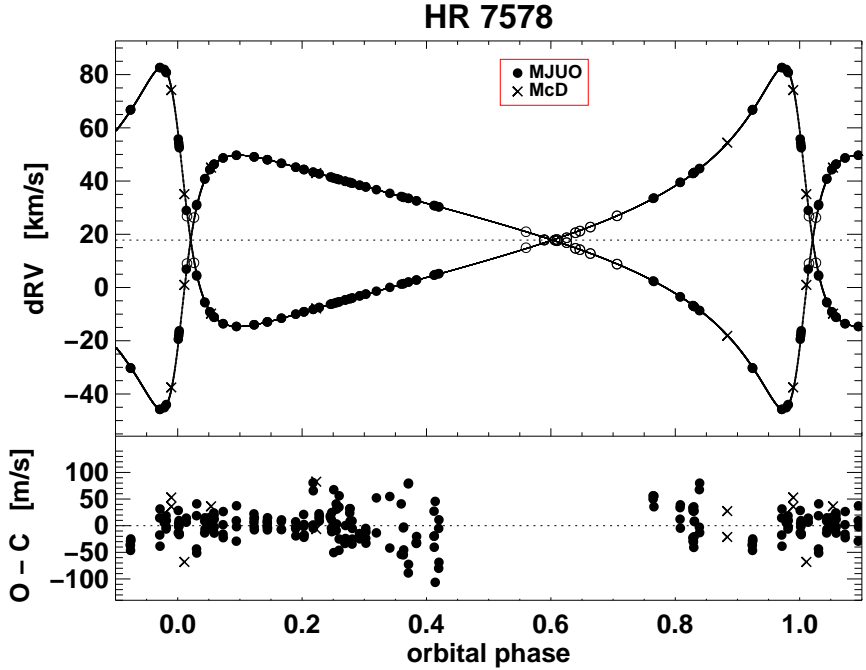}
	\caption{Radial velocity measurements (\textit{upper panel}) of HR7578 obtained with the modified Doppler code and their residuals (\textit{lower panel}). Filled dots represent MJUO spectra and crosses represent spectra taken at McDonald Observatory. Open circles represent the datapoints that were not used in the fit, as the spectral lines were too heavily blended, resulting in larger residuals. The solid line is our best-fit Keplerian orbit and the RMS around this model is $28.2\,\mathrm{m\,s^{-1}}$ for component A and $40.7\,\mathrm{m\,s^{-1}}$ for component B.}
	\label{fig:HD_188088_phased}
\end{figure}

\section{Summary and Discussion}
We have used the line index ratio of the $\mathrm{H}$-$\alpha$ and NaD lines to determine the amount of contamination in the spectra of $\alpha$ Cen A and B. Using the flux ratio from the line index ratio analysis and the binary orbit as an input to a modified Doppler code we can determine the true radial velocity of the main target for observations of $\alpha$ Cen A or B.

\citet{dumusque2012} estimated the effect of the contamination on the RV as a function of seeing and simply discarded observations that were too heavily affected. However, given that the vast majority of our spectra were contaminated to some extent (especially those of $\alpha$ Cen B), and the fact that it is crucial to have a very large number of observations, we developed a pipeline that applies the necessary corrections to the velocities. With the angular separation between $\alpha$ Cen A and B decreasing, this technique will sooner or later become necessary for all telescope sites.

After correcting for the contamination in a sample of 1307 observations of $\alpha$ Cen B, we have an RMS of typically $2.5\,\ms$ for the RV measurements of one night, compared to more than $100\,\ms$ for some heavily affected nights if no corrections were made. It is reassuring that the RMS of the corrected RVs is very similar to the RMS of the RVs we have obtained for single stars of similar spectral type.

The question whether or not we would be able to detect the proposed planet around $\alpha$ Cen B in our dataset has been addressed in a recent study by \citet{mike_astrobio}. After injecting a simulated planetary signal and pre-whitening the dataset by removing the binary orbit and the intrinsic stellar signals due to the long-term magnetic cycle and stellar rotational activity (similar to Dumusque et al.~2012 and Hatzes~2013), it was found that the simulated planetary signal emerges from the noise in the periodogram even if the white noise is as large as $\sim 5\,\ms$. The same study found that a super-Earth inside the HZ around $\alpha$ Cen B, which ranges from about 226 to 523 days in orbital period \citep[the narrow HZ as defined in][]{kaltenegger2013}, with a minimum mass of $3.2\,\mathrm{M_\oplus}$ and a period of 234 days should be detectable if the white noise does not exceed $\sim 3\,\ms$. We are therefore confident that the signal of the proposed 3.24-d planet should be detectable with our data if it exists, and we might also be able to detect a potential super-Earth in the habitable zone. Besides, recent studies have shown that planetary signals can be detected even if the signal is smaller than the overall RMS of the velocities \citep[e.g.][]{dumusque2012,tuomi2013,jenkins2014}.

In addition, we have successfully applied the new Doppler code to four double-lined spectroscopic binaries and have obtained velocities for both components simultaneously. The RV precision predictably depends on spectral type, but at present the precision of this method is mostly limited by the choice of the template stars. Using synthetic templates or more appropriate single stars as templates would almost certainly improve the RV precision still further. We have been able to improve the orbital parameters for these systems, in some cases substantially. Perhaps more importantly, the level of precision that can be obtained with this method will eventually allow us to search for planets in SB2 systems, which are usually excluded from RV planet search programmes, thus opening up new grounds for planet hunting.

To our knowledge the only other survey aimed at detecting planets in SB2 systems \citep{ratajczak2012} has not produced any discoveries. The potential discovery of a Jupiter-mass planet in a close triple system by \citet{konacki2005} could not be confirmed by \citet{eggenberger2007}. Eventually, this method could even be used to search for circumbinary planets as also proposed by \cite{konacki2009}.

\acknowledgments
\noindent \textit{Acknowledgements.} The work of CB has been supported by a University of Canterbury Doctoral Scholarship as well as by Marsden grant UOC1007, administered by the Royal Society of New Zealand. The Mt~John programme is also funded by Marsden grant UOC1007, administered by the Royal Society of New Zealand. This project is furthermore supported in part by the Australian Research Council Discovery Grant DP110101007. The authors would like to thank McDonald Observatory, University of Texas at Austin, for allowing the use of their Sandiford iodine cell at MJUO. We would also like to thank the referee N.~Haghighipour for his insightful comments, which helped improving this manuscript.


\begin{thebibliography}{}


\bibitem[Borucki et al., 2010]{kepler} 
  {Borucki}, W.~J. and {Koch}, D. and {Basri}, G. and {Batalha}, N. and 
	{Brown}, T. and {Caldwell}, D. and {Caldwell}, J. and {Christensen-Dalsgaard}, J. and 
	{Cochran}, W.~D. and {DeVore}, E. and {Dunham}, E.~W. and {Dupree}, A.~K. and 
	{Gautier}, T.~N. and {Geary}, J.~C. and {Gilliland}, R. and 
	{Gould}, A. and {Howell}, S.~B. and {Jenkins}, J.~M. and {Kondo}, Y. and 
	{Latham}, D.~W. and {Marcy}, G.~W. and {Meibom}, S. and {Kjeldsen}, H. and 
	{Lissauer}, J.~J. and {Monet}, D.~G. and {Morrison}, D. and 
	{Sasselov}, D. and {Tarter}, J. and {Boss}, A. and {Brownlee}, D. and 
	{Owen}, T. and {Buzasi}, D. and {Charbonneau}, D. and {Doyle}, L. and 
	{Fortney}, J. and {Ford}, E.~B. and {Holman}, M.~J. and {Seager}, S. and 
	{Steffen}, J.~H. and {Welsh}, W.~F. and {Rowe}, J. and {Anderson}, H. and 
	{Buchhave}, L. and {Ciardi}, D. and {Walkowicz}, L. and {Sherry}, W. and 
	{Horch}, E. and {Isaacson}, H. and {Everett}, M.~E. and {Fischer}, D. and 
	{Torres}, G. and {Johnson}, J.~A. and {Endl}, M. and {MacQueen}, P. and 
	{Bryson}, S.~T. and {Dotson}, J. and {Haas}, M. and {Kolodziejczak}, J. and 
	{Van Cleve}, J. and {Chandrasekaran}, H. and {Twicken}, J.~D. and 
	{Quintana}, E.~V. and {Clarke}, B.~D. and {Allen}, C. and {Li}, J. and 
	{Wu}, H. and {Tenenbaum}, P. and {Verner}, E. and {Bruhweiler}, F. and 
	{Barnes}, J. and {Prsa}, A. 2010, Science, 327, 977
\bibitem[Dumusque et al., 2012]{dumusque2012} 
  {Dumusque}, X. and {Pepe}, F. and {Lovis}, C. and {S{\'e}gransan}, D. and 
	{Sahlmann}, J. and {Benz}, W. and {Bouchy}, F. and {Mayor}, M. and 
	{Queloz}, D. and {Santos}, N. and {Udry}, S. 2012, Nature, 491, 207
\bibitem[Eggenberger et al., 2007]{eggenberger2007}
  {Eggenberger}, A. and {Udry}, S. and {Mazeh}, T. and {Segal}, Y. and 
	{Mayor}, M. 2007, A\&A, 466, 1179
\bibitem[Endl et al., 2000]{austral}
  {Endl}, M., {K\"urster}, M., \& {Els}, S.  2000, A\&A, 362, 585
\bibitem[Endl et al., 2014]{mike_astrobio}
  {Endl}, M. and {Bergmann}, C. and {Hearnshaw}, J. and {Barnes}, S.~I. and 
	{Wittenmyer}, R.~A. and {Ramm}, D. and {Kilmartin}, P. and {Gunn}, F. and 
	{Brogt}, E. 2014, IJA, this volume, in press
\bibitem[Fekel \& Beavers, 1983]{hd188088} 
  {Fekel}, F.~C., Jr., \& {Beavers}, W.~I. 1983, ApJ, 267, 682
\bibitem[Hatzes, 2013]{hatzes2013} 
  {Hatzes}, A.~P. 2013, ApJ, 770, 133
\bibitem[Hearnshaw et al., 2002]{HERCULES} 
  {Hearnshaw}, J.~B., {Barnes}, S.~I., {Kershaw}, G.~M., {Frost}, N., {Graham}, G., {Ritchie}, R., \& {Nankivell} G.~R. 2002, Experimental Astronomy, 13, 59 
\bibitem[Hearnshaw et al., 2013]{jbh_narit}
  {Hearnshaw}, J. and {Barnes}, S. and {Endl}, M. and {Wittenmyer}, R. 2013 in Planets in the Solar System and Beyond, 
	Proceedings of the $11^{\mathrm{th}}$ Asian-Pacific Regional IAU Meeting, National Astronomical Research Institute of Thailand
\bibitem[Jenkins \& Tuomi, 2014]{jenkins2014}
  {Jenkins}, J.~S. and {Tuomi}, M. 2014 arXiv1406.3093
\bibitem[Kaltenegger \& Haghighipour, 2013]{kaltenegger2013}
  {Kaltenegger}, L. and {Haghighipour}, N. 2013, ApJ, 777, 165
\bibitem[Konacki, 2005]{konacki2005}
  {Konacki}, M. 2005, Nature, 436, 230
\bibitem[Konacki et al., 2009]{konacki2009}
  {Konacki}, M. and {Muterspaugh}, M.~W. and {Kulkarni}, S.~R. and 
	{He{\l}miniak}, K.~G. 2009, ApJ, 704, 513
\bibitem[Mayor \& Queloz, 1995]{firstplanet} 
  {Mayor}, M. and {Queloz}, D. 1995, Nature, 378, 355
\bibitem[Mayor et al., 2003]{harps_mayor2003}
  {Mayor}, M. and {Pepe}, F. and {Queloz}, D. and {Bouchy}, F. and 
	{Rupprecht}, G. and {Lo Curto}, G. and {Avila}, G. and {Benz}, W. and 
	{Bertaux}, J.-L. and {Bonfils}, X. and {Dall}, T. and {Dekker}, H. and 
	{Delabre}, B. and {Eckert}, W. and {Fleury}, M. and {Gilliotte}, A. and 
	{Gojak}, D. and {Guzman}, J.~C. and {Kohler}, D. and {Lizon}, J.-L. and 
	{Longinotti}, A. and {Lovis}, C. and {Megevand}, D. and {Pasquini}, L. and 
	{Reyes}, J. and {Sivan}, J.-P. and {Sosnowska}, D. and {Soto}, R. and 
	{Udry}, S. and {van Kesteren}, A. and {Weber}, L. and {Weilenmann}, U. 2003, The Messenger, 114, 20
\bibitem[Pourbaix et al., 2002]{orbit}
  {Pourbaix}, D., {Nidever}, D., {McCarthy}, C., {Butler}, R.~P., {Tinney}, C.~G., {Marcy}, G.~W., {Jones}, H.~R.~A., {Penny}, A.~J., {Carter},
	B.~D., {Bouchy}, F., {Pepe}, F., {Hearnshaw}, J.~B., {Skuljan}, J., {Ramm}, D., \& {Kent}, D. 2002, A\&A, 386, 280
\bibitem[Quintana et al., 2014]{kepler186} 
  {Quintana}, E.~V. and {Barclay}, T. and {Raymond}, S.~N. and 
	{Rowe}, J.~F. and {Bolmont}, E. and {Caldwell}, D.~A. and {Howell}, S.~B. and 
	{Kane}, S.~R. and {Huber}, D. and {Crepp}, J.~R. and {Lissauer}, J.~J. and 
	{Ciardi}, D.~R. and {Coughlin}, J.~L. and {Everett}, M.~E. and 
	{Henze}, C.~E. and {Horch}, E. and {Isaacson}, H. and {Ford}, E.~B. and 
	{Adams}, F.~C. and {Still}, M. and {Hunter}, R.~C. and {Quarles}, B. and 
	{Selsis}, F. 2014, Science, 344, 277
\bibitem[Ratajczak et al., 2012]{ratajczak2012}
  {Ratajczak}, M. and {Konacki}, M. and {Kulkarni}, S.~R. and 
	{Muterspaugh}, M.~W. 2012 in From Interacting Binaries to Exoplanets: Essential Modeling Tools,
	Proceedings of the IAU Symposium S282, Cambridge University Press
\bibitem[Tuomi et al., 2013]{tuomi2013}
  {Tuomi}, M. and {Jones}, H.~R.~A. and {Jenkins}, J.~S. and {Tinney}, C.~G. and 
	{Butler}, R.~P. and {Vogt}, S.~S. and {Barnes}, J.~R. and {Wittenmyer}, R.~A. and 
	{O'Toole}, S. and {Horner}, J. and {Bailey}, J. and {Carter}, B.~D. and 
	{Wright}, D.~J. and {Salter}, G.~S. and {Pinfield}, D. 2013, 551, A79
\bibitem[Wang et al., 2012]{boottran}
  {Wang}, Sharon, X. and {Wright}, J.~T. and {Cochran}, W. and 
	{Kane}, S.~R. and {Henry}, G.~W. and {Payne}, M.~J. and {Endl}, M. and 
	{MacQueen}, P.~J. and {Valenti}, J.~A. and {Antoci}, V. and 
	{Dragomir}, D. and {Matthews}, J.~M. and {Howard}, A.~W. and 
	{Marcy}, G.~W. and {Isaacson}, H. and {Ford}, E.~B. and {Mahadevan}, S. and 
	{von Braun}, K. 2012, ApJ, 761, 46
\bibitem[Wittenmyer et al., 2014]{rob2014}
  {Wittenmyer}, R.~A. and {Endl}, M. and {Bergmann}, C. and {Hearnshaw}, J. and 
	{Barnes}, S.~I. and {Wright}, D. 2014 in Formation, Detection, and Characterization of Extrasolar Habitable Planets, Proceedings of the IAU Symposium 293,
	Cambridge University Press
\bibitem[Wright \& Howard, 2009]{rvlin}
  {Wright}, J.~T. and {Howard}, A.~W. 2009, ApJS, 182, 205
\bibitem[Wright \& Howard, 2013]{rvlin_erratum}
  {Wright}, J.~T. and {Howard}, A.~W. 2013, ApJS, 205, 22

  
\end{thebibliography}
\end{document}